\documentclass[preprint,double-spaced,aps,apssymb]{revtex4-1}
\usepackage{amsmath,amssymb}
\usepackage{graphicx}
\usepackage{dcolumn}
\usepackage{bm}
\usepackage{subfigure}
\newcommand{\ber}{\begin{eqnarray}}
\newcommand{\eer}{\end{eqnarray}}
\newcommand{\bea}{\begin{equation}}
\newcommand{\eea}{\end{equation}}

\begin{document}

\title{Directed transport in equilibrium : analysis of the dimer model with inertial terms}

\author{A. Bhattacharyay}
\email{a.bhattacharyay@iiserpune.ac.in}
 \affiliation{Indian Institute of Science Education and Research, Pune, India}

\date{\today}

\begin{abstract}
We have previously shown an analysis of our dimer model in the over-damped regime to show directed transport in equilibrium. Here we analyze the full model with inertial terms present to establish the same result. First we derive the Fokker-Planck equation for the system following a Galilean transformation to show that a uniformly translating equilibrium distribution is possible. Then, we find out the velocity selection for the centre of mass motion using that distribution on our model. We suggest generalization of our calculations for soft collision potentials and indicate to interesting situation with possibility of oscillatory non-equilibrium state within equilibrium.
\end{abstract}
\maketitle
In a previous communication \cite{ari} we have shown a simplified calculation at the over-damped limit of a dimer model with particles undergoing hard collisions. With this model, we demonstrated the possibility of achieving sustained broken symmetry for the system of dimer in contact with a heat bath in the absence of any external drive. As a result of that sustained broken symmetry under equilibrium conditions, the dimer system shows directed transport. We have already given a lengthy discussion in ref.\cite{ari} on why equilibrium does not forbid possibility of directed uniform motion, how the model be generalized for soft collisions and the possible role of dissipation at collision, when the dimer-bath system as a whole remains isolated from surroundings, causing uniform directed motion in equilibrium. In this paper, we do a detailed calculation treating the full model (containing inertial terms) to show that our previous results at over-damped limit was correct. Our present results, being identical to the over-damped ones we have already shown, will establish that the hardcore collisions we had considered in the previous case \cite{ari} through boundary conditions was a correct way of dealing with it at the over-damped limit.
\par
Our dimer model \cite{ari} with hardcore collisions between particles of the dimer would look like the following.
\ber\nonumber
\dot{v_1} = \lim_{\epsilon \rightarrow 0}\Theta(Z-\epsilon)[&-&(1-\beta)v_1+\sqrt{2(1-\beta)k_BT}\eta_1(t)]\\ &-& \lim_{\epsilon \rightarrow 0}\frac{\partial}{\partial z}\left[ \Theta ( z-\epsilon)\frac{\alpha z^2}{2} + (1-\Theta ( z-\epsilon))\left( a-\frac{az}{\epsilon} \right) \right ]\\\nonumber
\dot{v_2} = \lim_{\epsilon \rightarrow 0}\Theta(Z-\epsilon)[&-&v_2+\sqrt{2k_BT}\eta_2(t)]\\ &+& \lim_{\epsilon \rightarrow 0}\frac{\partial}{\partial z}\left[ \Theta ( z-\epsilon)\frac{\alpha z^2}{2} + (1-\Theta ( z-\epsilon))\left( a-\frac{az}{\epsilon} \right) \right ]\\\nonumber
\dot{z} = v_1 - v_2
\eer
In the above model $z=x_1-x_2$ where $x_1$ and $x_2$ are the positions of the particles of the dimer such that $x_1 \geq x_2$. As we have shown in the previous article \cite{ari} that, the requirement of having two different damping during the particles of the dimer are on flight (i.e. when the harmonic force is working on them) and during the period they physically collide with each other is the basis of the sustained symmetry breaking. Here, since we are going to analyze the system with hardcore collisions, for the sake of simplicity, we consider that the time over which actual collision takes place is negligibly small. For example, the time of collision is of the order of time of a single collision of the dimer particles with a bath molecule (say). Within this consideration, we can safely drop the noise and the damping during collision and we also drop the harmonic interaction during collision because this force is proportional to the separation of particles which is practically zero during collision. we emphasize on the fact that, taking this hardcore limit is for the sake of simplicity and one can readily generalize this model for soft-core collisions where one must take a different damping constant during the time the collision happens and the strength of the noise in accordance with this damping during collision to have the symmetry breaking. Having the broken symmetry there, one would find the qualitative result (i.e. directed transport in equilibrium) is present, of course, with some more variety. We would talk about this generalization in the end of this article. Note that, $<\eta_i(t)>=0$ and $<\eta_i(t)\eta_j(t)>=\delta_{ij}\delta(t_1-t_2)$. The mass of the particles are taken the same and unity for the sake of simplicity which would make the numerical simulation of the above model implementing hardcore collisions straightforward. In the above model, consider that the parameter $a$ in the collision potential is so large that the particles have no chance to pass through each other for a given $k_BT$.
\par
Following a Galilean transformation to a moving frame moving with a velocity $V_{cm}$ one can rewrite the above model in the form
\ber\nonumber
\dot{v_1} = \lim_{\epsilon \rightarrow 0}\Theta(Z-\epsilon)[&-&(1-\beta)(v_1-V_{cm})+\sqrt{2(1-\beta)k_BT}\eta_1(t)]\\ &-& \lim_{\epsilon \rightarrow 0}\frac{\partial}{\partial z}\left[ \Theta ( z-\epsilon)\frac{\alpha z^2}{2} + (1-\Theta ( z-\epsilon))\left( a-\frac{az}{\epsilon} \right) \right ]\\\nonumber
\dot{v_2} = \lim_{\epsilon \rightarrow 0}\Theta(Z-\epsilon)[&-&(v_2-V_{cm})+\sqrt{2k_BT}\eta_2(t)]\\ &+& \lim_{\epsilon \rightarrow 0}\frac{\partial}{\partial z}\left[ \Theta ( z-\epsilon)\frac{\alpha z^2}{2} + (1-\Theta ( z-\epsilon))\left( a-\frac{az}{\epsilon} \right) \right ]\\
\dot{z} = v_1 - v_2 .
\eer
Considering the time dependent probability of a state ($z,v_1,v_2$) being given as \cite{swa}
\bea
P(z,v_1,v_2,t)=<\delta(z-z(t))\delta(v_1-v_1(t))\delta(v_2-v_2(t))>,
\eea
where $<>$ means average over noise. In general, for the Gaussian noise $\eta$, the probability distribution of the noise is given by $P[\eta] = C\exp(-\int\limits_{t_0}^{t_f}{dt\frac{\eta(t)^2}{2\Gamma k_BT}})$, where $C$ is normalization constant and $\Gamma$ being the damping is to be taken $(1-\beta)$ for $\eta_1$ and unity for $\eta_2$ in our actual calculations. Its generally taken that the time gap $t_f-t_0 = \tau \gg \tau_c$, where $\tau_c$ is the correlation time for noise which is practically zero for the $\delta$-correlated noise approximation. The noise average of a quantity $X[\eta]$ is obtained by evaluating the functional integral $\int {\cal D}[\eta]X[\eta]P[\eta]$, where the discrete time steps are $\tau$. Since, in our hardcore collision limit, we are considering the collision time is of the order of $\tau_c$, it would not practically affect the calculations and a very standard result would follow as one does the relevant averaging over the noise. Taking into consideration the continuity equation for the normalized probability $P(z,v_1,v_2,t)$, we can easily derive the Fokker-Planck (FP) equation for the above mentioned model on a co-moving frame from
\ber\nonumber
\frac{\partial P}{\partial t} &=& -\frac{\partial }{\partial z}<\delta(z-z(t))\delta(v_1-v_1(t))\delta(v_2-v_2(t))(v_1-v_2)>\\\nonumber &&- \frac{\partial}{\partial v_1}<\delta(z-z(t))\delta(v_1-v_1(t))\delta(v_2-v_2(t))\\\nonumber &&[ \lim_{\epsilon \rightarrow 0}\Theta(Z-\epsilon)[-(1-\beta)(v_1-V_{cm})+\sqrt{2(1-\beta)k_BT}\eta_1(t)]\\\nonumber && - \lim_{\epsilon \rightarrow 0}\frac{\partial}{\partial z}[ \Theta ( z-\epsilon)\frac{\alpha z^2}{2} + (1-\Theta ( z-\epsilon))( a-\frac{az}{\epsilon})]]>\\\nonumber && -\frac{\partial}{\partial v_2}<\delta(z-z(t))\delta(v_1-v_1(t))\delta(v_2-v_2(t))\\\nonumber && [\lim_{\epsilon \rightarrow 0}\Theta(Z-\epsilon)[-(v_2-V_{cm})+\sqrt{2k_BT}\eta_2(t)]\\ &&+ \lim_{\epsilon \rightarrow 0}\frac{\partial}{\partial z}[ \Theta ( z-\epsilon)\frac{\alpha z^2}{2} + (1-\Theta ( z-\epsilon)( a-\frac{az}{\epsilon}) ]]>
\eer
The FP equation that follows would now look like
\ber\nonumber
\frac{\partial P}{\partial t} &+& (v_1-v_2)\frac{\partial P}{\partial z} - \frac{\partial \Phi(z)}{\partial Z}\left( \frac{\partial P}{\partial v_1} - \frac{\partial P}{\partial v_2} \right )\\\nonumber &=& \lim_{\epsilon \rightarrow 0}\Theta(Z-\epsilon) (1-\beta)\frac{\partial}{\partial v_1}\left( (v_1-V_{cm})P+k_BT\frac{\partial P}{\partial v_1}\right)\\ &+& \lim_{\epsilon \rightarrow 0}\Theta(Z-\epsilon)\frac{\partial}{\partial v_2}\left( (v_2-V_{cm})P+k_BT\frac{\partial P}{\partial v_2}\right ).
\eer
The form of the above mentioned FP equation is independent of the choice of the internal field potential $\Phi(z)$. So, long as the field is internal, the FP equation can be written in the above mentioned form because one can always write down the model in terms of the internal coordinates and the individual particle velocities. Since the system would settle in equilibrium at the minimum of the potential (of course with fluctuations present), $\Phi(z)$ being internal the system and the field can move together and the above mentioned form of the FP equation simply mentions this fact without giving a selection for the velocity of CM which has to be found out from the dynamics.
\par
The steady state or equilibrium solution of the above FP equation is the standard one but with a constant drift and can be written down as
\bea
P(Z,v_1,v_2)=Ne^{-\Phi(z)/k_BT}e^{-(v_1-V_{cm})^2/2k_BT}e^{-(v_2-V_{cm})^2/2k_BT}
\eea
where $N$ is the normalization constant and $\Phi(z)=\lim_{\epsilon \rightarrow 0}[\Theta ( z-\epsilon)\frac{\alpha z^2}{2} + (1-\Theta ( z-\epsilon))( a-\frac{az}{\epsilon})]$.

\par
having found the equilibrium solution of the relevant FP equation to our model, we would now use this distribution to find the various average quantities to get a selection of the uniform average drift $V_{cm}$. To do that, we would first write down our model into the form involving $V_{cm}= (v_1+v_2)/2$ and $V_z=v_1-v_2$. So, now the model looks like
\ber
\dot{V}_{cm} &=& \lim_{\epsilon \rightarrow 0}\Theta(Z-\epsilon)[(\frac{\beta}{2}-1)V_{cm}+\frac{\beta}{4}V_z+\xi_{cm}]\\\nonumber
\dot{V}_z&=& \lim_{\epsilon \rightarrow 0}\Theta(Z-\epsilon)[(\frac{\beta}{2}-1)V_z+\beta V_{cm}+\xi_{z}-2\alpha z]-2\lim_{\epsilon \rightarrow 0}\frac{\alpha z^2}{2}\delta(z-\epsilon)\\ &&- 2\lim_{\epsilon \rightarrow 0}\left[ \frac{a}{\epsilon}(1-\Theta(z-\epsilon))+(a-\frac{az}{\epsilon})\delta(z-\epsilon))\right ]\\
\dot{z}&=& V_z
\eer
In the above form of the model $\xi_{cm}=(\sqrt{2(1-\beta)k_BT}\eta_1(t)+\sqrt{2k_BT}\eta_2(t))/2$ and $\xi_z=\sqrt{2(1-\beta)k_BT}\eta_1(t)-\sqrt{2k_BT}\eta_2(t)$. Its easy to see that $<\xi_{cm}>=<\xi_z>=0$, but, $\xi_{cm}$ and $\xi_z$ would have cross correlations and that is why we have taken the other form of the model involving $v_1$ and $v_2$ to derive the FP equation. Now the analysis is straight forward, replace for $\Theta(z-\epsilon)V_z$ in Eq.11 from Eq.10 then take the average with respect to the above mentioned probability distribution. The $<V_{cm}>$ now comes out the constant $V_{cm}$ which is the uniform drift of the frame. Set all the average accelerations equal to zero (equilibrium) and averages of the $\xi_i$ equal to zero. The contribution from the collision part would vanish as one takes the limit because in the first collision term $<\Theta(z-\epsilon)>=1$ and in the second collision term $z$ is replaced by $\epsilon$ due to the presence of the delta function. Upon simplification the velocity of the centre of mass would come out as
\bea
V_{cm} = -\frac{\beta\alpha}{2(1-\beta)}<z>
\eea
which is exactly the same as the one we had got in the over-damped calculations. Note that finding $<V_z>$ is simple in the present case because we know the velocity distributions of $v_1$ and $v_2$ and $<V_z>=<v_1-v_2>$ can be checked to be vanishing. Note that $<V_z>$ cannot be found out from the Eq.10 because it does not take into account the part of $V_z$ that comes during the collision happens and that is why $<\Theta(z-\epsilon)V_z>$ has a non-vanishing value whereas $<V_z>$ be vanishing in equilibrium. Its not difficult to see that, had we taken the collision contribution to the $V_z$ considering the presence of damping during collision, the collision terms contributing to this part of $V_z$ would have a factor $1/(1-\Theta(z-\epsilon))$ instead of $1/\Theta(z-\epsilon)$ as is there for $V_{cm}$. Now, the collision part would become non-vanishing unlike the case of its contribution to the $V_{cm}$. Before we go for a generalization of our model to the soft collision case, let us first compare our result with that obtained from the direct numerical simulation of the full model comprising of Eq.1-3. 
\par
Fig.1 shows four plots of the $\ln(|V_{cm}|)$ against $\ln(\alpha)$ at four different values of $k_BT$. All these plots are straight lines with slope almost equal to 0.5. The match between the theoretical and numerical values are quite good. In the previous communication \cite{ari} where we had mentioned that there is some order of magnitude mismatch between the theoretical and numerical value of $V_{cm}$ that was because of not taking proper scaling of the noise term with time interval. In the over-damped case, $\bigtriangleup x = \sqrt{2k_BT/\Gamma}\sqrt{\bigtriangleup t}$ where $\Gamma $ is the damping) is basically the statement of Fick's Law of diffusion. So, any numerical value put in the simulation for the $k_BT$ can be directly taken for analytic calculations. But, for the full (containing inertial term) model simulation we use $\bigtriangleup u = \sqrt{2\Gamma k_BT}\sqrt{\bigtriangleup t}$ where $u$ is velocity. We have to convert this expression to the equivalent form of the Fick's Law and then calculate the $k_BT$ and that would make $k_BT = \bigtriangleup t^3n$ where we have put the number $n$ in the place of $k_BT$ in the numerical analysis. In this rescaling $\bigtriangleup t^2$ comes from the conversion of $\bigtriangleup u$ to $\bigtriangleup x$ and the rest comes because of the gamma appearing in the numerator of the noise term mentioned above. Note that in the numerical simulation to implement hardcore collision we simply exchange the position and velocity of the particles following any overshooting and that we can do because of considering the same mass for the particles. The time step in our simulation has been taken $\bigtriangleup t = 0.01$.
\begin{figure}
{\includegraphics[width=10 cm,angle=-90]{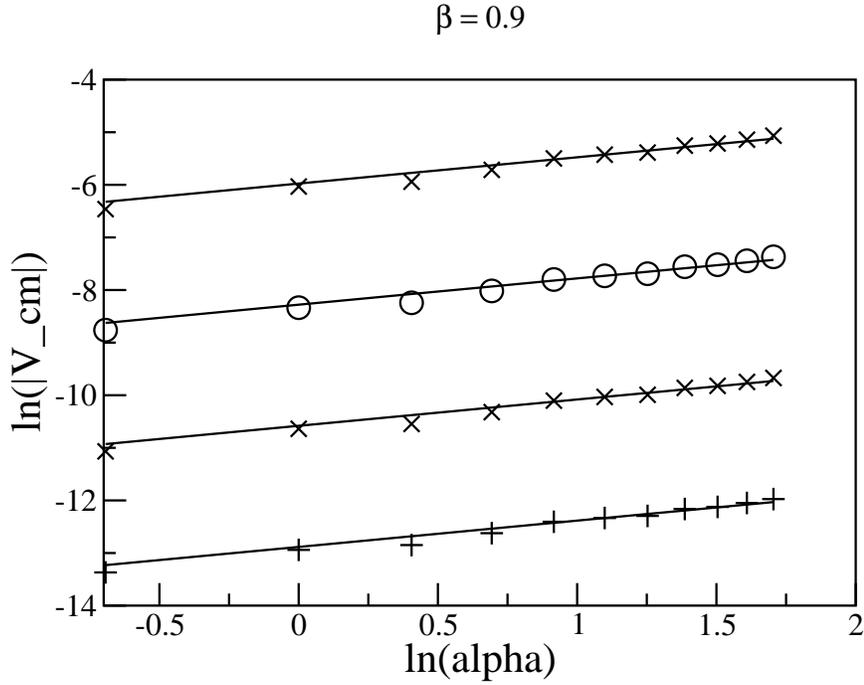}}
\caption [Figure 1]{Plot of $\ln(|v_{cm}|)$ vs $\ln(\alpha)$ at $\beta = 0.9$ for values of $k_BT=5\times10^{-7}$, $k_BT=5\times10^{-9}$, $k_BT=5\times10^{-11}$, and $k_BT=5\times10^{-13}$ respectively from top to bottom. The straight lines are plots of theoretical values and symbols correspond to numerical results.}
\end{figure}

\par
Having understood the model with hardcore collisions, let us generalize it for soft collisions. At the core of the symmetry breaking in the model is the consideration of two different damping, one during collision and the other when the particles are not in touch with each other but are interacting harmonically. Let us consider that the particles have a damping $\Gamma^\prime$ when they collide which is different from that they have on flight (namely $(1-\beta)$ and 1). This may happen because when particles are actually in touch with each other their surroundings are different from that when they are on flight. This $\Gamma^\prime$ can also be different for different particles of the dimer, but, for the sake of simplicity one may consider them to be equal and that would not change the qualitative result. Now considering the collision potential is soft and of width $\epsilon$ one can rewrite the model as
\ber\nonumber
\dot{v_1} = \Theta(Z-\epsilon)[&-&(1-\beta)v_1+\sqrt{2(1-\beta)k_BT}\eta_1(t)]\\ &-& \frac{\partial}{\partial z}\left[ \Theta ( z-\epsilon)\frac{\alpha z^2}{2} + (1-\Theta ( z-\epsilon))\left( a-\frac{az}{\epsilon} \right) \right ]\\\nonumber &&-(1-\Theta ( z-\epsilon))[\Gamma^\prime v_1 - \sqrt{2\Gamma^\prime k_BT}\eta_1(t)]\\\nonumber
\dot{v_2} = \Theta(Z-\epsilon)[&-&v_2+\sqrt{2k_BT}\eta_2(t)]\\ &+& \frac{\partial}{\partial z}\left[ \Theta ( z-\epsilon)\frac{\alpha z^2}{2} + (1-\Theta ( z-\epsilon))\left( a-\frac{az}{\epsilon} \right) \right ]\\\nonumber &&-(1-\Theta ( z-\epsilon))[\Gamma^\prime v_2 - \sqrt{2\Gamma^\prime k_BT}\eta_2(t)]\\\nonumber
\dot{z} = v_1 - v_2 .
\eer
Notice that, the temperature remains the same ($T$) irrespective of collision or harmonic interaction is taking place. This model can be analyzed similarly as the one shown above and would result in a centre of mass motion of the system over a range of values of $\Gamma^\prime$ and $\beta$ even showing conditions for the directional reversal of $V_{cm}$. We will show a detailed analytic and numerical treatment of the above model in future. Imagine the fact that, the dimer and the bath composite system is thermally insulated from the surroundings. Now, also imagine that at collision some other internal degrees of freedom of the particles are excited (particles are not dimensionless particles) and that results in an inelastic collision of the particles. Eventually, these internal coordinates would relax leaving the energy back to the bath and consider that this process happens pretty quickly. Now, this isolated system must reach equilibrium over time since its an isolated system from the surroundings, but, intermittent stealing of energy by the internal degrees of freedoms of the particles would result in a kinetic energy oscillation of the dimer. This kinetic energy oscillation can be imagined as an oscillation of the effective temperature of the particles at shorter time scales smaller than the one over which the average temperature for the equilibrium is defined. One can even tune the time period of this effective temperature oscillation of the dimer by tuning the force constant $\alpha$ because it sets the time period of oscillation of the dimer on average. Not only that, this time period can also be large. This is a typical so-called non-equilibrium state (on average steady oscillation) which would be coexisting with equilibrium because the isolated system must reach equilibrium at large time. So, this would be a possible way of getting a non-equilibrium order within the time scale over which equilibrium is defined and its not all uncorrelated fluctuations within the equilibrium time scale. There will be associated positive and negative productions of entropy  (rather entropy exchange between bath and the particles) such that the total entropy production rate over a cycle remains zero to maintain equilibrium at larger time scales. So, this system is an entropy oscillator which is consistent with equilibrium defined over a somewhat larger scale.

\end{document}